\documentclass[aps,prb,twocolumn,showpacs,floats]{revtex4-1}
\usepackage[dvips]{graphicx}
\usepackage{amsmath}
\usepackage{amssymb}
\usepackage{empheq}
\newcommand{\eps}{\epsilon}

\newcommand{\ra}{\rangle}
\newcommand{\la}{\langle}
\newcommand{\ud}{\mathrm{d}}

\begin{document}

\title{\bf Frequency dependent transport through a spin chain}

\author{Kevin A. van Hoogdalem and Daniel Loss}

\affiliation{Department   of  Physics,   University   of  Basel,
  Klingelbergstrasse      82,     CH-4056      Basel,     Switzerland}

\date{\today}

\begin{abstract}
Motivated by potential applications in spintronics, we study frequency dependent spin transport in nonitinerant one-dimensional spin chains. We propose a system that behaves as a capacitor for the spin degree of freedom. It consists of a spin chain with two impurities a distance $d$ apart. We find that at low energy (frequency) the impurities flow to strong coupling, thereby effectively cutting the chain into three parts, with the middle island containing a discrete number of spin excitations. At finite frequency spin transport through the system increases. We find a strong dependence of the finite frequency characteristics both on the anisotropy of the spin chain and the applied magnetic field. We propose a method to measure the finite-frequency conductance in this system.
\end{abstract}

\pacs{75.10.Jm, 75.10.Pq, 75.30.Ds, 75.76.+j}
\maketitle

\section{Introduction}
One of the main issues in modern electronics is that as devices get ever smaller the removal of waste energy generated by Joule heating (which is in turn caused by the scattering of the information carrying conduction electrons) becomes problematic. A possible solution to this problem of excess heating is offered by spintronics~\cite{Aws07,Zut04,Aws02} in nonitinerant materials. Since charge transport is absent in such materials, the energy that is dissipated in the transport of a single information carrier (a spin) in a nonitinerant system~\cite{Trauz08} is much lower than in both itinerant electronic- and spintronic devices.~\cite{Hal06,Ovc08} Furthermore, it has been theoretically shown that the energy required for a single logic operation is low in a spin based device.~\cite{Che09}

Clearly, energy considerations are not the only factor that play a role in the determination of the feasibility of spintronics in nonitinerant materials. Other important criteria are the operating speed of potential devices, and the possibility of creation, control, and read out of the signals used in these systems. Indeed, it has been proposed that spin-based devices in nonitinerant systems could operate at high frequencies.~\cite{Aws07} Furthermore, there have been various proposals to create (by means of spin pumping~\cite{Bra02} and the spin Seebeck effect~\cite{Uch10}) and read out (by means of the inverse spin Hall effect~\cite{Kaj10,San11}) these pure spin currents in nonitinerant systems. Logic schemes that are based on interference of spin waves in insulating magnets have been proposed.~\cite{Kos05,Sch08} In these schemes, the dispersion relation of the spin waves is controlled either by magnetic fields, magnetic textures,~\cite{Dug05} electric fields due to the induced Aharonov-Casher phase~\cite{Mei03} or in multiferroic materials,~\cite{Mil08,Rov10} or spin orbit coupling.~\cite{Liu11}

In this work we focus on transport of spin excitations in antiferromagnetic nonitinerant one-dimensional quantum systems (spin chains). In such systems charge transport is absent, and the elementary excitations that transport magnetization are spin waves. Of the possible candidates for such systems that exist we specifically mention here Cs$_2$CoCl$_4$, which has a large anisotropy in the exchange interaction,~\cite{Ken02} and SrCuO$_2$, in which transport was recently shown to be ballistic over distances on the order of 1 $\mu$m (see Ref. \onlinecite{Hlu10}). 

Inspired by possible applications in spintronics, we set out to design the spin analogues in insulating magnets of the different components that are used in modern electronics. Recently, we have shown that systems containing spin chains can be used as a spin resistance~\cite{Mei03} and a spin diode.~\cite{Hoo11} In this work, we propose a device based on spin chains that mimics the working of a capacitor, but for the spin degree of freedom. That is, a device that has a spin conductance that is zero for applied DC driving fields and increases with frequency for applied AC driving fields. Traditional capacitors have a wide range of applications in electronics, most notably in frequency selection, noise reduction, and temporary energy storage.

The device that we propose consists of a spin chain with anisotropy $\Delta_{ij}$ in the exchange interaction in the $z$ direction. We introduce two impurities, that each model the replacement of a single atom, in the spin chain. A renormalization group argument shows that the impurities flow to strong coupling at low energies, or, equivalently, low frequencies. At low frequencies the chain is therefor in an insulating state, with an island containing a discrete number of spin excitations in the middle. We show that at higher frequency the spin conductance becomes nonzero; the system undergoes a transition to a conducting state. We show that the point at which this transition occurs can be tuned by an external magnetic field and the bulk anisotropy in the exchange interaction. Transport through such double barrier systems has received attention recently,~\cite{Fur93,Sas95,Pec03,Rec06,Wan10} but these studies were restricted to the itinerant electronic case. Here we focus on the nonitinerant magnetic case with an emphasis on applications.

In Section \ref{sec:2} we introduce our system and the techniques we use to analyze the system. Explicit expressions for the finite frequency spin conductance of the system under consideration are derived in Section \ref{sec:3}. This finite frequency spin conductance contains all the required information to show that our system behaves as a spin capacitor. In Section \ref{sec:4} we present numerical results, conclusions are given in Section \ref{sec:5}.

\section{System and model}
\label{sec:2}
\begin{figure}
\centering
\includegraphics[width=0.9\columnwidth]{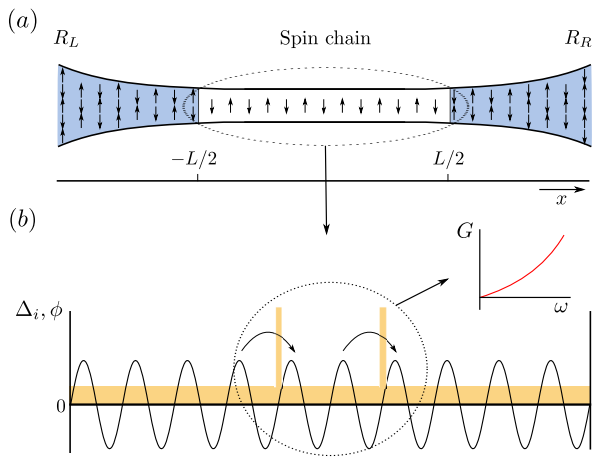}
\caption[]{(a) Schematic view of the nonitinerant spin system. The 1-D spin chain of finite length $L$ is adiabatically connected (meaning that the length of the transition region $L_t \gg 4 a$)  to two spin reservoirs. (b) Illustration of the working principle behind the spin capacitor. Strong impurities pin the field $\phi(\vec{r})$, thereby blocking the system for spin transport.}
\label{fig:rect}
\end{figure}
A spin chain of length $L$ (see Fig. \ref{fig:rect}), adiabatically connected to two three-dimensional spin reservoirs, can be described by the anisotropic XXZ-Heisenberg Hamiltonian
\begin{equation}
H = \sum_{\la ij\ra} J_{ij}\left[S_i^xS_{j}^x + S_i^yS_{j}^y + \Delta_{ij} S_i^zS_{j}^z\right] + g \mu_B \sum_i B_i S_i^z.
\end{equation}
The summation is over nearest neighbors, and we avoid double counting of the bonds. Here $S_i^{\alpha}$ is the $\alpha$-component of the spin operator at position $\vec{r}_i$. $J_{ij}$ denotes the exchange interactions between the two nearest neighbors at $\vec{r}_i$ and $\vec{r}_j$. We assume an, in general spatially varying, anisotropy $\Delta_{ij} > 0$ in the exchange coupling. This anisotropy typically originates either from a Dzyaloshinksii-Moriya (DM) or dipole-dipole interaction between the different spins.~\cite{Nol09} The last term in the Hamiltonian describes the Zeeman coupling between the external time-dependent magnetic field, $\vec{B}_i(t) = B_i(t) \hat{e}_z$, and the spin operator. We assume the magnetic field to be of the form $B_i(t) = B_0  -\left[1- \theta(-L/2)\right] \Delta B \cos \omega t$.

For our spin capacitor, we need to introduce two impurities in the spin chain. We propose to use an impurity of a specific type, a single instance of which, located at $\vec{r}_{i_0}$, is described by its anisotropy
\begin{equation}
 \Delta_{ij} = \Delta \delta_{i,j-1}\left(\delta_{i,i_0}+ \delta_{i,i_0+1}\right) \equiv \Delta_{i_0}.
\label{eq:imp1}
\end{equation}
Such an impurity describes the replacement of a single ion in the spin chain by another ion with a different anisotropy. We choose this form because it contains all the important physics for a general impurity. With two impurities, located at $\vec{r}_i = \pm d/2$, and allowing for a constant anisotropy $\Delta'$ (which we assume to be smaller than 1), the anisotropy becomes
\begin{equation}
 \Delta_{ij} = \Delta_{-d/2} + \Delta_{d/2} + \Delta'.
\label{eq:Anis}
\end{equation}

It is well known~\cite{Gia03} that in the one-dimensional spin chain we can map the Heisenberg Hamiltonian on a fermionic Hamiltonian by means of the Jordan-Wigner transformation. Assuming that $g \mu_B B \ll J$ and $\Delta_{ij} \ll 1$ we can take the continuum limit in the resulting fermionic Hamiltonian, which we can then linearize around the Fermi wave vector. In the process we need to introduce left- and right going fermions $\psi^\dagger_{L/R}(\vec{r})$. We can then perform a bosonization procedure~\cite{Del98,Sen} using the density field $\phi(\vec{r})$ and its conjugate field $\theta(\vec{r})$, which satisfy $\left[ \phi(x),\partial_{x'}\theta(x')\right] = i\pi \delta(x-x')$. Using the fermionic creation operator for $r = L/R$-going particles
\begin{eqnarray}
\psi^{\dagger}_r(x) &=& \frac{1}{\sqrt{2\pi a}}e^{-i\eps_rk_Fx}e^{i\left[\eps_r\phi(x)-\theta(x)\right]},
\label{eq:fercrea}
\end{eqnarray}
where $a$ is the distance between two neighboring spins, and $\eps_r = \pm 1$, we arrive at the following real-time action describing the system
\begin{equation}
 S[\phi] = S_0[\phi] + S_z[\phi] + S_B[\phi].
\end{equation}
Here $S_0[\phi]$ is the quadratic action which originates from the in-plane terms and the forward scattering terms in the out-of-plane part of the Heisenberg Hamiltonian. It is given by
\begin{eqnarray}
S_0[\phi] &=& \frac{\hbar}{2\pi K} \int \ud^2 r \left[ u \left(\partial_x \phi(\vec{r})\right)^2-\frac{1}{u}\left(\partial_t \phi(\vec{r})\right)^2\right].
\label{eq:QuadHam}
\end{eqnarray}
Here $u$ is the propagation velocity of the bosonic excitations and $K$ is the Luttinger Liquid interaction parameter which can be determined from the parameters $J$ and $\Delta'$ of the spin chain from the Bethe-Ansatz.~\cite{Hal80} We denote $\vec{r} = (x,t)$. $S_B[\phi]$ describes the Zeeman coupling between magnetic field and the $z$ component of the spin operator 
\begin{equation}
S_{B}[\phi]  =  - \frac{g \mu_B }{\pi} \int \ud^2 r \phi(\vec{r}) \partial_x B(\vec{r}).
\end{equation}
Lastly, $S_z[\phi]$ contains the most relevant contributions coming from the two impurities. We have shown previously that, for an impurity of the form Eq. (\ref{eq:imp1}) centered at $x_0$, the most relevant terms are given by~\cite{Hoo11}
\begin{eqnarray}
S_{x_0}[\phi] & = & \frac{\sigma}{\pi^2} \int \ud t \left[ \lambda^a_{x_0}\cos 2\phi(x_0,t) + \lambda^b_{x_0}\sin2\phi(x_0,t)\right].
\label{eq:SI}
\end{eqnarray}
Here, the couplings $\lambda_{x_0}^{a/b}$ are of the order $J\Delta$. Furthermore, $\sigma = K g\mu_B B_0 /(\hbar \omega_c)$ is the doping away from half-filling due to the magnetic field $B_0$. $\omega_c$ is the UV-cutoff of the theory, which is on the order of $J/\hbar$. We have thus $S_z[\phi] = S_{-d/2}[\phi]+S_{d/2}[\phi]$. Note that we must assume here that $2k_F d \neq \pi$, since otherwise the two impurities would cancel each other. To lowest order the backscattering terms flow under renormalization as $1-K$. Since we work in the regime $1/2 \leq K < 1$ (which corresponds to $ 0 < \Delta' \leq 1$) the impurities flow to strong coupling for low energies. In this regime, the spin chain is effectively cut into three parts. The middle island ($-d/2 \leq x \leq d/2)$ contains a discrete number of excitations, and tunneling through the island is only possible by co-tunneling processes, in which two excitations hop on and off the middle island simultaneously. These processes are strongly suppressed at low temperatures. For finite frequency, processes similar to photon-assisted tunneling become possible, so that we expect the spin current through the spin chain to increase as a function of frequency. 

From Eq. (\ref{eq:SI}) it is seen that the strength of the impurities is proportional to the filling $\sigma$. Hence we can control the finite frequency characteristics of the system by applying a constant magnetic field $B_0$ to our system. Most strikingly, this field allows us to control the frequency at which the system will become conducting in the spin sector. Since the Fermi wave vector also depends on $B_0$, this field also controls the number of spins that are on the island for a given width $d$ of the island. The number of spins relative to half-filling can be estimated as $(d/a)\sigma$.

Lastly, we can model the excitations in the two reservoirs by the quadratic Luttinger Liquid action as well. We will assume two reservoirs that are described by the the Heisenberg Hamiltonian with $\Delta = 0$, so that the excitations in the reservoirs are free, and we have $K_r = 1$.

\section{Finite-frequency spin conductance}
\label{sec:3}
As long as we consider only magnetic fields that point along the quantization axis, the spin current $I_s(\vec{r})$ through the system can be determined from the conservation of spin, and is given by
\begin{equation}
 I_s(\vec{r}) = -\frac{g\mu_B}{\pi} \partial_t \phi(\vec{r}).
\end{equation}
The nonlocal differential finite frequency conductivity $\sigma(x,x',\omega)$ is defined as the Fourier transform $\sigma(x,x',\omega) = \int \ud t e^{i\omega t} \sigma(x,x',t)$ of the real-time linear response of the spin current to an infinitesimal change in the time-dependent magnetic field gradient
\begin{equation}
 \sigma(x,x',t-t') = \frac{\delta \left\la I_s(\vec{r})\right\ra}{\delta [\partial_{x'}B(\vec{r}')]}.
\end{equation}
In circuit theory, the quantity which is of interest is the spin conductance $G(\omega)$, which we define for our system as 
\begin{equation}
G(\omega) = \textrm{Re}\left[\sigma(-L/2,L/2,\omega)\right].
\end{equation}
With this convention a positive magnetic field difference $\Delta B$ yields a positive spin current for the free system. To calculate the transport properties we use the Keldysh formalism.~\cite{Kel65} We assume that at $t = -\infty$ the system is described by the quadratic action $S_0[\phi] + S_B[\phi]$ and that the perturbation $S_z[\phi]$ is turned on adiabatically. We calculate the conductivity to lowest non-zero order in $S_z[\phi]$. In this description the conductivity is given by
\begin{equation}
\sigma(x,x',t-t') = i\frac{g\mu_B}{\pi} \frac{\delta}{\delta [\partial_{x'}B(\vec{r}')]}\left.\left(\frac{\delta Z[J]}{\delta J(\vec{r})}\right)\right|_{J=0}.
\label{eq:part}
\end{equation}
Here, $Z[J]$ is the partition function of the system [see Eq. (\ref{eq:partfun})], and $J(\vec{r})$ a generating functional. We give a short explanation of the evaluation of Eq. (\ref{eq:part}) in Appendix \ref{ch:app1} and summarize the important results here. We can show that the conductivity consists of two contributions
\begin{equation}
 \sigma(x,x',\omega) = \sigma_0(x,x',\omega) + \sigma_{\textrm{BS}}(x,x',\omega).
\end{equation}
The first contribution to the conductivity describes the system in the absence of impurities. It gives the exact spin current for the free system, since in the absence of impurities the current is linear in magnetic field gradient. The free conductivity is given by
\begin{equation}
 \sigma_0(x,x',\omega) = -\left(\frac{g\mu_B}{\pi \sqrt{\hbar}}\right)^2 \left(i\omega\right) i G_0^R(x,x',\omega).
\end{equation}
The retarded Green's function for $x,x'$ in the chain or at the boundaries can be obtained from analytic continuation of the Matsubara Green's function for the inhomogeneous Luttinger Liquid model.~\cite{Mas95} If we assume infinitesimal dissipation in the leads, the retarded Green's function is given by
\begin{eqnarray}
 G_0^R(x,x',\omega) = \frac{\pi K}{2\omega} \bigg\{ 1 +  \frac{1}{\gamma^{-1}e^{-i\tilde{\omega}} - \gamma e^{i\tilde{\omega}}}\times \nonumber\\
\sum_{r=\pm 1} \bigg[ e^{ir\tilde{\omega}(\tilde{x}+\tilde{x}')} + \gamma^r \left(e^{-ir\tilde{\omega}|\tilde{x}-\tilde{x}'|}+r\right)e^{ir\tilde{\omega}}\bigg]\bigg\}.
\end{eqnarray}
Here $\gamma = (1-\kappa)/(1+\kappa)$ is the reflection coefficient for Andreev-like (in the sense that for some processes the transmission coefficient can be larger than 1) density wave scattering at the chain-reservoir interface, and $\kappa = K / K_r$. The dimensionless frequency $\tilde{\omega} = \omega/\omega_L$, where $\omega_L = u / L$ is the energy scale determined by the finite length of the spin chain. Also $\tilde{x} = x/L$. In the limit $\tilde{\omega} \to 0$ we have $\sigma(x,x',\omega) = K_r(g\mu_B)^2/h$ as expected.

The quantity $\sigma_{\textrm{BS}}(x,x',\omega)$ contains corrections to the conductivity coming from the impurities. For the situation with two impurities, it depends on the amplitude and both the absolute and relative position of the impurities, and is given by 
\begin{equation}
 \sigma_{\textrm{BS}}(x,x',\omega) = \left(\frac{\pi \sqrt{\hbar}}{g\mu_B}\right)^4 \sum_{\xi,\xi' = \pm d/2} \sigma_{\xi,\xi'} (x,x',\omega).
\label{eq:sigmaBS}
\end{equation}
As it turns out, the terms with $\xi = \xi'$ reduce to the known result for a single impurity,~\cite{Dolc05,Saf08} the terms with $\xi \neq \xi'$ come from interference between the impurities located at $-d/2$ and $d/2$. Here $\sigma_{\xi,\xi'} (x,x',\omega)$ is given by
\begin{eqnarray}
 \sigma_{\xi,\xi'}(x,x',\omega) & = & - \frac{\sigma_0(x,\xi,\omega)}{\hbar\omega} \big[\sigma_0(\xi',x',\omega)iK(\xi,\xi',\omega) + \nonumber\\
& &  -\sigma_0(\xi,x',\omega)iK(\xi,\xi',0)\big].
\end{eqnarray}
$K(\xi,\xi',\omega)$ denotes the retarded backscattering current-backscattering current correlator, which is defined by
\begin{equation}
iK(\xi,\xi',\omega) = \int_0^\infty \ud \tau e^{i\omega \tau}\left\la \left[I_{\textrm{BS}}(\xi,\tau),I_{\textrm{BS}}(\xi',0)\right]\right\ra.
\label{eq:K}
\end{equation}
The backscattering current operator is defined as
\begin{eqnarray}
 I_{\textrm{BS}}(x_0,t) & = & \frac{g \mu_B}{\pi \hbar} \frac{\delta S_{x_0}[\phi]}{\delta \phi(x_0,t)}\nonumber\\
 & = & 2 \frac{g\mu_B}{\pi \hbar} \left[ \lambda_{x_0}^b \cos 2 \phi(x_0,t) -\lambda_{x_0}^a \sin 2 \phi(x_0,t) \right].
\end{eqnarray}
In taking the functional derivative, $\phi(x_0,t)$ is regarded as a function of $t$ only. Some lengthy but straightforward algebra shows that we can rewrite Eq. (\ref{eq:K}) as
\begin{eqnarray}
 iK(\xi,\xi',\omega) & = & 2\left(\frac{g\mu_B}{\pi \hbar}\right)^2\left[\lambda_\xi^a\lambda_{\xi'}^a + \lambda_\xi^b\lambda_{\xi'}^b\right]\int_0^\infty \ud \tau e^{i\omega\tau}\nonumber\\
& & \times\left[e^{2C_0(\xi,\xi',\tau)}-e^{2C_0(\xi',\xi,-\tau)}\right],
\label{eq:K2}
\end{eqnarray}
where $C_0(x,x',\tau)$ is the regularized $\phi\phi$-correlator at zero temperature for the free Luttinger Liquid action on a system of finite length $L$, given by
\begin{eqnarray}
 &&C_0(x,x',\tau)  = \nonumber\\
&&\frac{K}{2}\bigg\{ \sum_{n \in \textrm{even}}\gamma^{|n|} \ln \left[\frac{n^2+\alpha^2}{(n+\tilde{x} - \tilde{x}')^2+\alpha^2(i\bar{\tau}+1)^2}\right] + \nonumber\\
&&\frac{1}{2}\sum_{n\in \textrm{odd}} \gamma^{|n|}\ln\left[\frac{[(n+2\tilde{x})^2+\alpha^2][(n+2\tilde{x}')^2+\alpha^2]}{\left[(n+\tilde{x}+\tilde{x}')^2+\alpha^2(i\bar{\tau}+1)^2\right]^2}\right]\bigg\}.
\end{eqnarray}
Here $\bar{\tau} = \omega_c \tau$, $\alpha = \omega_L/\omega_c$, and $\omega_c$ is the UV-cutoff of the theory, which is on the order of $J/\hbar$. Finally, we have to replace the bare couplings $\lambda$ in Eq. (\ref{eq:K2}) by the renormalized couplings, given by $(\max\left[\omega_L,\omega\right]/\omega_c)^{-1+K}\lambda$. In the next Section we will evaluate Eq. (\ref{eq:sigmaBS}) numerically.
\section{Results}
\label{sec:4}
\begin{figure}
\centering
\includegraphics[width=0.99\columnwidth]{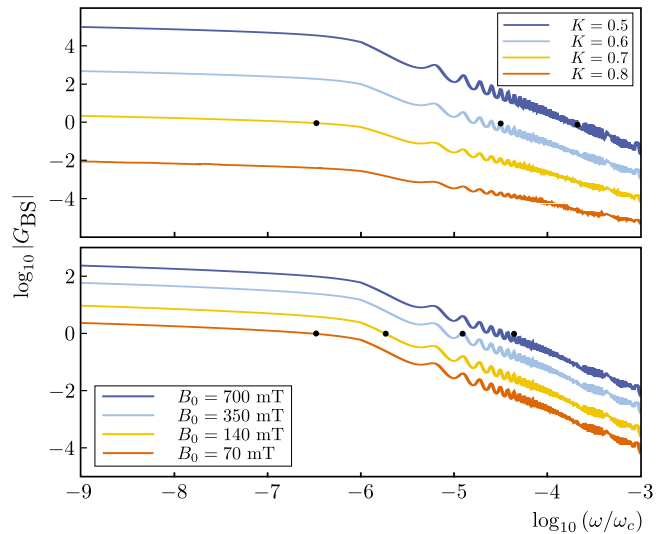}
\caption[]{(top) Plot of $\log_{10} |G_{\textrm{BS}}(\omega)|$ (in units of $(g\mu_B)^2/h$) as a function of $\log_{10} (\omega/\omega_c)$ for different values of the Luttinger liquid interaction parameter $K$ (increasing from top to bottom). Used parameters are $B_0 = 70$ mT, $J = 10^2$ K, $L = 10^6 a$, $d = 10^4 a$, $K_r = 1$, and $\Delta = 0.3$. The black dots denote the frequencies $\omega^*$ where $|G_0(\omega^*)| = |G_{\textrm{BS}}(\omega^*)|$. Our perturbative results cannot be used for $\omega < \omega^*$. (bottom) Idem for different values of $B_0$ (decreasing from top to bottom). Used parameters are $K = 0.7$, $J = 10^2$ K, $L = 10^6 a$, $d=10^4a$, $K_r = 1$, and $\Delta = 0.3$. The oscillatory behavior is explained in the caption of Fig. \ref{fig:3} and the associated part in the text.}
\label{fig:2}
\end{figure}
We have shown in Sec. \ref{sec:3} that the spin conductance $G(\omega) = \textrm{Re}\left[\sigma(-L/2,L/2,\omega)\right]$ of our system consists of two parts, the free conductance $G_0(\omega)$ and the backscattering conductance $G_{\textrm{BS}}(\omega)$. These are of opposite sign for low frequencies. Fig. \ref{fig:2}a shows a plot of $\log_{10} |G_{\textrm{BS}}(\omega)|$ ($G_{\textrm{BS}}(\omega)$ is here in units of $(g\mu_B)^2/h$) as a function of $\log_{10} (\omega/\omega_c)$ for different values of the Luttinger liquid parameter $K$. Different values of $K$ correspond to different values of the constant anisotropy $\Delta'$ in the spin chain (see Eq. (\ref{eq:Anis})). It is seen that $G_{\textrm{BS}}(\omega)$ goes to zero at high frequencies. In the high-frequency regime the spin conductance is therefore determined by the free conductance $G_0(\omega)$ only, and hence the system is conducting. For lower frequencies the backscattering conductance $G_{\textrm{BS}}(\omega)$ grows, so that the total spin conductance decreases. Since our calculation of $G_{\textrm{BS}}(\omega)$ is perturbative in the strength of the impurities, we cannot trust our calculations at frequencies $\omega \lesssim \omega^*$, where $\omega^*$ is the frequency such that $|G_{\textrm{BS}}(\omega^*)| = |G_0(\omega^*)|$. Indeed, instead of the behavior shown in Fig. \ref{fig:2} the total spin conductance $G(\omega)$ is approximately zero at frequencies below $\omega^*$. Therefor, $\omega^*$ is approximately the frequency at which the system undergoes the transition from insulating to conducting behavior.

Interestingly, we see that the backscattering conductance is strongly dependent on the Luttinger liquid parameter $K$. If we focus again on the conductance at the frequency $\omega^*$, we see that an increase from $K = 0.6$ to $K = 0.7$ leads to a decrease of the frequency at which the system switches between conducting and insulating behavior of approximately 2 orders of magnitude. We note that we have $\omega_c \approx J/\hbar$, which for $J = 10^2$ K is approximately $\omega_c \approx 2.1 \cdot 10^{12}$ s$^{-1}$. The previously discussed increase in $\omega^*$ is therefore approximately from $7.4*10^5$ rad s$^{-1}$ to $6.7*10^7$ rad s$^{-1}$.

We see from Fig. \ref{fig:2} that the frequency $\omega^*$ strongly depends on the applied magnetic field $B_0$ as well. This is caused by the fact that the strength of the impurity is proportional to $B_0$, see Eq. (\ref{eq:SI}). This strong dependence is convenient for possible applications, since it would allow for an externally tunable capacitor. For the parameters in Fig. \ref{fig:2} we see that by increasing $B_0$ from $B_0 = 70$ mT to $B_0 = 700$ mT, we can increase $\omega^*$ from $7.4 *10^5$ rad s$^{-1}$ to $9.5 *10^7$ rad s$^{-1}$.

\begin{figure}
\centering
\includegraphics[width=0.99\columnwidth]{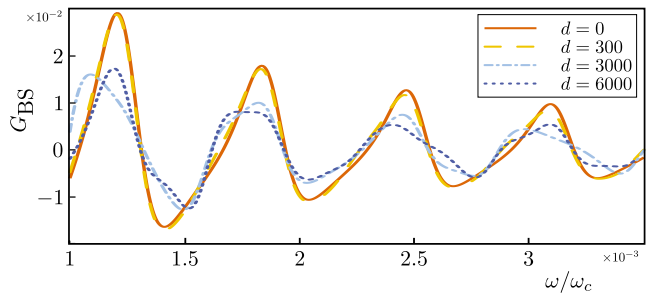}
\caption[]{Backscattering conductance $G_{\textrm{BS}}(\omega)$ as a function of $\omega/\omega_c$ for different values of the impurity separation $d$ (in units of $a$). Used values are $B_0 = 1.4$ T, $J = 10^2$ K, $L = 10^4 a$, $K=0.7$, $K_r = 1$, and $\Delta = 0.2$. It is seen that, independent of $d$, the conductance oscillates with period $2 \pi \omega_L$ (we estimate $\omega_L \approx a/L$ here, the values for the free system). The values of $d$ only influences the shape of the oscillations, not the period.}
\label{fig:3}
\end{figure}

In Fig. \ref{fig:3} we show the dependence of $G_{\textrm{BS}}(\omega)$ on the distance $d$ between the two impurities. Regardless of $d$, it is seen that the conductance oscillates with frequency $2\pi \omega_L$ (see caption). These oscillations are caused by interference of the bosonic excitations due to Andreev-like reflections at the boundaries between reservoir and spin chain. It has been shown that, in the presence of a double barrier structure in the chain, additional oscillations with frequency determined by the parameters $d/L$ and $(L-d)/(2L)$ are visible in the current at finite driving field.~\cite{Wan10} These oscillations do not show up in the conductance at zero $\Delta B$; however, they are expected show up in the conductance at finite $\Delta B$. Here, $d$ changes the shape of the oscillations, but not the frequency. All the calculations here were done at $T = 0$. Generally, the effect of finite temperature is to wash out the fluctuations.

The finite frequency spin conductance could be measured by measuring the spin accumulation in a system consisting of a spin diode of the form proposed in Ref. \onlinecite{Hoo11} in series with the spin capacitor proposed here, between two spin reservoirs. We can apply the driving field $\Delta B(t)$ to the left reservoir, and measure the spin accumulation in the right reservoir. In this setup, the spin accumulation is zero when the spin capacitor is in the insulating state. As was shown in Ref. \onlinecite{Hoo11}, if we consider $10^4$ parallel spin chains with $K=0.6$ in a magnetic field $B_0 = 750$ mT, and amplitude of the driving field $\Delta B = 43$ mT, we have a spin accumulation of approximately $10^{12}$ magnons per seconds if the capacitor is in the conducting state.

\section{Conclusions}
\label{sec:5}
In this work we have proposed a spin capacitor in a system consisting of one-dimensional nonitinerant spin chain adiabatically connected to two spin reservoirs. The spin chain is required to have an anisotropy in the exchange interaction. We have shown that the replacement of a single atom in the spin chain leads to an local backscattering term in the Hamiltonian that flows to strong coupling at low energies. By including two such impurities in the spin chain we got a device which is insulating at zero frequency driving field $\Delta B$, and has a spin conductance that grows in magnitude under an increase in frequency of the driving field. We have studied the influence of the anisotropy in the exchange interaction $\Delta$ and the external magnetic field $B_0$, and have found that both have a strong influence on the finite frequency characteristics of the system. We have proposed a way to measure the effects.
\label{conclusion}
\section{Acknowledgements}
This work has been supported by the Swiss NSF and NCCR Nanoscience Basel.
\appendix
\section{Partition function}
\label{ch:app1}
In this Section we discuss the partition function of the system described in Sec. \ref{sec:2}. For simplicity, we consider a system that is described by the action $S_0[\phi] + S_B[\phi] + \frac{\sigma\lambda_0^a}{\pi^2}\int_{-\infty}^{\infty} \ud t \cos \left[2 \phi(0,t)\right]$. The partition function of the system described in Sec. \ref{sec:2} can be inferred from the results for this action in a straightforward manner. The partition function can be written in vectorized form as
\begin{eqnarray}
Z[J(\vec{r})] &=& \int D \vec{\phi} \exp \bigg[ -\frac{1}{2} \int \ud^2 r \ud^2 r' \times\nonumber\\
 &&\left[ \vec{\phi}(\vec{r})\cdot G_0(\vec{r},\vec{r}')\vec{\phi}(\vec{r}') - 2i\delta(\vec{r}-\vec{r}')\vec{\phi}(\vec{r})\cdot Q \vec{J}(\vec{r}')\right]\bigg] \times\nonumber\\
 &&\exp \left[\frac{i\sigma\lambda_0^a}{\pi^2\hbar}\int_{-\infty}^{\infty} \ud t \sum_{\eta = \pm} \eta \cos \left[2 \phi^{\eta}(0,t)\right]\right],
\label{eq:partfun}
\end{eqnarray}
where we have defined
\begin{equation}
\begin{split}
\vec{\phi}(\vec{r}) & =  \left(\begin{array}{c} \phi^+(\vec{r}) \\ \phi^-(\vec{r}) \end{array}\right), \\
\vec{J}(\vec{r}) &= \sqrt{2}\left(\begin{array}{c} -\frac{g \mu_B}{\pi \hbar}\partial_x B(\vec{r}) \\ J(\vec{r})/2 \end{array}\right), \\
Q &= \frac{1}{\sqrt{2}} \left(\begin{array}{cc} 1 & -1 \\ 1 & 1 \end{array}\right), \\
G_0(\vec{r},\vec{r}') & =  \left(\begin{array}{cc} G_0^{++}(\vec{r},\vec{r}') & G_0^{+-}(\vec{r},\vec{r}') \\ G_0^{-+}(\vec{r},\vec{r}') & G_0^{--}(\vec{r},\vec{r}') \end{array}\right). \end{split}
\end{equation}
Here $G_0^{\eta,\eta'}(\vec{r},\vec{r}')$ is the contour-ordered Green's function
\begin{equation}
G_0^{\eta,\eta'}(\vec{r},\vec{r}') = \left\la T_c \phi^{\eta}(\vec{r})\phi^{\eta'}(\vec{r}')\right\ra_{S_0}.
\end{equation}
Here, $\eta,\eta' = \pm$ denotes whether a field is located on the positive or negative Keldysh contour. Eq. (\ref{eq:part}) can be evaluated by first making the linear transformation $\vec{\phi}'(\vec{r}) = \vec{\phi}(\vec{r}) - i \int \ud \vec{r}'G_0(\vec{r},\vec{r}') Q^T \vec{J}(\vec{r}')$ and then performing the functional derivatives.~\cite{Dolc05}

\end{document}